\documentclass[rnote]{aa} 
\usepackage{graphicx}
\usepackage{epstopdf}
\usepackage{natbib}
\usepackage{txfonts}
\begin{document}
   \title{Near infrared and gamma-ray monitoring of TANAMI gamma-ray bright sources}

   \author{R. Nesci\inst{1}, 
             G. Tosti\inst{2},
             T. Pursimo\inst{3},
             R. Ojha\inst{4,5},
             M. Kadler\inst{6,7},
          }

   \offprints{R. Nesci}

   \institute{INAF-IAPS, via Fosso del Cavaliere 100, 00133, Roma Italy, and Universita' La Sapienza, Roma, Italy \\
              \email{roberto.nesci@uniroma1.it}\\
            \and
            Universita' di Perugia, 06023 Perugia, Italy \\
             \and
            Nordic Optical Telescope, Apartado 474, E-38700 Santa Cruz de La Palma,
            Santa Cruz de Tenerife, Spain \\
            \and
            NASA Goddard Space Flight Center, Astrophysics Science Division, Code~661, Greenbelt, MD 20771, USA\\  
            \and 
            Institute for Astrophysics \& Computational Sciences, Catholic University  of America, Washington, DC 20064, USA\\
            \and
            Institut f\"ur Theoretische Physik und Astrophysik, Lehrstuhl f\"ur Astronomie, Univ.\ W\"urzburg, Emil-Fischer           Str.\ 31, 97074 W\"urzburg, Germany\\
           \and
           Dr.\ Karl-Remeis-Sternwarte  and Erlangen Centre for Astroparticle Physics, Sternwartstra\ss{}e~7, 96049 Bamberg, Germany
             }

   \date{Received 13 January 2013 / Accepted 7 May 2013}

 
  \abstract
  {Spectral energy distribution and its variability are basic tools for understanding the physical processes operating in active galactic nuclei (AGN).}
  {In this paper we report the results of a one-year near infra red (NIR)  and optical monitoring of a sample of 22 AGN known to  be gamma-ray emitters, aimed at discovering correlations between optical and gamma-ray emission.}
  {We observed our objects with the Rapid Eye Mount (REM) telescope in J, H, K, and R bands nearly twice every month during their visibility window and derived light curves and spectral indexes. We also analyzed the gamma-ray data from the \textit{Fermi} gamma-ray Space Telescope, making weekly averages.}
   {Six sources were never detected during our monitoring, proving to be fainter than their historical Two micron all sky survey (2MASS) level. All of the sixteen detected sources showed marked flux density variability, while the spectral indexes remained unchanged within our sensitivity limits. Steeper sources showed, on average, a larger variability.  From the NIR light curves we also computed a variability speed index for each detected source.   
Only one source (PKS\,0208-512) underwent an NIR flare during our monitoring. Half of the sources showed a regular flux density trend on a one-year time scale, but do not show any other peculiar characteristic. The broadband spectral index $\alpha_{ro}$ appears to be a good proxy of the NIR spectral index only for BL Lac objects. 
No clear correlation between NIR and gamma-ray data is evident in our data, save for PKS 0537$-$441, PKS 0521$-$360, PKS 2155$-$304, and PKS 1424$-$418.
The gamma-ray/NIR flux ratio showed a large spread, QSO being generally gamma-louder than BL Lac, with a marked correlation with the estimated peak frequency ($\nu_{peak}$) of the synchrotron emission. }
   {}

  \authorrunning{R. Nesci et al. } 
  \titlerunning {NIR-Gamma monitoring ...}

   \maketitle
%

\section{Introduction}
\label{sect:intro}
The spectral energy distribution of a radio loud active galactic nucleus (AGN) can be described as a double bell shape in a log($\nu$)-log($\nu F_{\nu}$) plane. The low-frequency bump, in the radio to infrared/optical/X-ray band, is mainly produced by synchrotron emission of relativistic electrons  streaming down the jet \citep{Blandford1979}. The high-frequency bump, peaking in the MeV to TeV band, probably arises from inverse Compton scattering processes \citep{Konigl1981,  Sikora2001} although hadronic processes are also a possibility \citep{Mannheim1995}. According to the position of the peak frequency of the synchrotron emission $\nu_{s}$, the sources are classified as low, intermediate or high Synchrotron peaked (LSP, ISP, or HSP); LSP  are typically radio-selected and HSP are typically X-ray selected  \citep{Abdo2010}.

The study of the temporal evolution of the emission from synchrotron and inverse Compton processes can give strong constraints to the physical models trying to explain the structure of the emitting regions and the relative importance of the different processes. For this purpose we selected a subsample of the TANAMI AGN sample \citep{Ojha2010}, currently monitored at radio frequencies both at single dish and interferometric mode, to be observed in the optical-near infrared bands (R$_C$, J, H, and K) during the year 2011. We looked for possible flare activity and correlation between the gamma-ray flux, monitored by the \textit{Fermi} Large Area Telescope (LAT) and the optical-NIR flux.
Our subsample was selected on the basis of the J-band apparent magnitude of the sources, as given by the 2MASS catalogue \citep{Cutri2003}: only objects brighter than J=16.5 were selected, in order to be monitored with the Rapid Eye Mount (REM) telescope \citep{Chincarini2003}. REM is  a 60\,cm robotic instrument located in the southern hemisphere at La Silla Observatory and operated by INAF\footnote {http://www.rem.inaf.it}. 

In the Roma Blazar Catalogue (BZCat\footnote{www.asdc.asi.it/bzcat/}) \citep{Massaro2009}, 18 of our sources are classified as QSO, BL Lac, or of uncertain classification (U), based on optical spectroscopy; as not all objects were listed in this catalogue,  we looked at the classification reported on the TANAMI web page\footnote{http://pulsar.sternwarte.uni-erlangen.de/tanami/}. For the 18 sources in common we found 7 discrepant classifications.  We finally  decided to adopt the more recent classification reported in the \textit{Fermi} second LAT AGN catalogue (2LAC) \citep{Ackermann2011} for all the sources.

The list of the targets is given in Table~\ref{table:1}, where we report in Col. 1 the historical name, in Col. 2 the Second Fermi Gamma-ray LAT catalogue (2FGL) identification, in Col. 3 the classification in the 2LAC, in Col. 4 the redshift (z); in Col. 5 the SED shape classification (LSP, ISP or HSP) in the 2LAC; in Col. 6 our SED shape estimate (see Sect. 2); in Col. 7 a quality flag (see Sect. 2); the subsequent columns will be described in Sect. 2.


\section{NIR and optical observations}
\label{sect:niroptical}
An average of 12 optical-NIR observations for each source were made during the campaign, with uneven sampling mainly because of seasonal constraints. Photometry of the sources was made using IRAF/Apphot\footnote{IRAF is distributed by the National Optical Astronomical Observatory, which is operated by AURA, under contract with National Science Foundation.}. Aperture photometry was performed with a 2 arcsec radius and several nearby stars, generally encompassing the source, were taken as references from the 2MASS catalogue \citep{Cutri2003} to transform instrumental magnitudes into the 2MASS JHK photometric system. 
For each frame a linear fit was computed between instrumental and catalogue magnitude of the stars, providing a best-fit slope generally very close to one and from this fit the source magnitude was then derived. Internal photometric errors were derived from Apphot, using the REM infrared camera parameters (electrons to counts conversion factor and readout noise). A more realistic estimate is provided by the root mean square deviation of the reference stars from the best-fit line. The best signal to noise ratio was usually obtained in the J-band.

The R-band images were often of poor quality, mainly as a result of poor focusing (after our runs, the optical camera was replaced with a new one in August 2012). Therefore, we have a limited number of useful R-band images and these are of the brightest sources; these magnitudes proved to be of little help and will not be considered.

From the NIR observations, our sample of 22 sources was divided into three groups:
\begin{enumerate}
\item  sources with photometric errors in all three IR bands small enough to allow a meaningful color index measurement (below $\sim$0.05 mag, 7 sources);
\item sources bright enough to be measured but with larger errors, so that only a J-band light curve can reliably be obtained (9 sources);
\item sources not detected or unreliable sources (6 sources).
\end {enumerate}

Group membership is listed in Col. 7 of Table~\ref{table:1}. In Group 1 there are  five BL Lac, one unclassified (PKS\, 0521$-$365) and a radio galaxy (PKS\, 0625$-$354); in Group 2 there are four BL Lac and five QSO; in Group 3 there are three BL Lac, two QSO and one uncertain. Group 3 sources will not be considered in this paper.

The classification of PKS\, 0625$-$354 as a radio galaxy is still debated; we will return to it in Sect. 3.
\begin{table*}
\caption{Source sample}             
\label{table:1}      
\centering                          
\begin{tabular}{lllllllllllllll}        
\hline\hline                 
Name &            2FGL &        2LAC &  z   & SED & SED & Gr & mean-J & range & sigma &J(2M) & tr & N&  $\alpha$REM & $\alpha$2M \\  
         &                       &        class&       & 2LAC& our   &       &              &           &            &          &    &   &                         &                     \\
\hline                        
PKS\,0047$-$579 & 0049.7-5738 & BZQ  & 1.797 &  - & -   & 3 &  -   &     -     &  -   &   -   &   - &  - &   -   &   - \\
PKS\,0208$-$512 & 0209.5-5229 & BZB  & 0.999 &  - &ISP  & 2 &14.89 & 13.7-16.0 & 0.83 & 13.69 &  no &  - & -     & 1.14\\
PKS\,0332$-$403 & 0334.2-4008 & BZB  & 1.445 &  - &  -  & 3 &  -   &     -     &  -   &   -   &   - &  - &   -   &   - \\
PMN\,J0334$-$3725& 0334.3-3728& BZB  &       &LSP & ISP & 2 &14.52 & 13.8-15.4 & 0.45 & 14.85 & dec & -  & -     & 1.23\\
PKS\,0402$-$362 & 0403.9-3604 & BZQ  & 1.417 & LSP& -   & 3 &  -   &     -     &  -   &   -   &   - &  - &   -   &   - \\  
PKS\,0447$-$439 & 0449.4-4350 & BZB  & 0.107 & -  & HSP & 1 &12.68 & 12.2-13.1 & 0.39 & 13.89 &  inc & 17 & 0.47 & 0.57 \\
PKS\,0454$-$463 & 0456.1-4613 & BZQ  & 0.852 & -  & LSP&  2 &15.80 & 15.2-16.4 & 0.39 & 15.38 & dec  & -  &  -   & 0.50 \\
PKS\,0521$-$365 & 0523.0-3628 & AGN  & 0.055 & ISP& LSP & 1 &12.87 & 12.5-13.1 & 0.20 & 12.95 &   no & 12 & 0.95 & 0.94 \\
PKS\,0537$-$441 & 0538.8-4405 & BZB  & 0.894 & -  & LSP & 1 &13.21 & 12.5-14.0 & 0.42 & 13.23 &   no & 14 & 0.98 & 1.16 \\
PKS\,0625$-$354 & 0627.1-3528 & RG   & 0.054 & LSP& LSP & 1 &13.11 & 12.8-13.3 & 0.18 & 13.38 &   no & 11 & 0.15 & 0.46 \\
PKS\,0637$-$752 & 0635.5-7516 & BZQ  & 0.653 & -  & LSP&  2 &14.52 & 14.0-15.0 & 0.26 & 14.85 &  no  & -  &  -   & 0.03:\\
PKS\,0700$-$661 & 0700.3-6611 & BZB  &       & -  &ISP &  2 &14.18 & 14.0-14.6 & 0.18 & 13.60 & no   & -  &      & 0.65 \\
PMN\,J0810$-$7530&0811.1-7527 & BZB  &       & -  & - &   3 &  -   &     -     &  -   &   -   &   -  &  - &   -  &   -  \\
PKS\,1144$-$379 &1146.8-3812  & BZQ  & 1.048 & LSP& LSP & 2 &15.70 & 13.6-15.5 & 0.44 & 15.66 &  no  & -  &  -   & 1.12 \\
PMN\,J1329$-$5608& 1329.2-5608& AGU  &       & -  & -   & 3 &  -   &     -     &  -   &   -   &   -  &  - &   -  &   - \\
PKS\,1424$-$418 &1428.0-4206  & BZQ  & 1.522 &  - & LSP & 2 &14.52 & 13.6-15.5 & 0.49 & 16.26 &  inc &  - &  -   & 1.07 \\
PMN\,J1444$-$3908& 1443.9-3908& BZB  & 0.065 & HSP& HSP & 1 &13.62 & 13.2-14.0 & 0.28 & 13.66 &  wav & 10 & -0.02 & 0.18 \\
PMN\,J1603$-$4904& 1603.8-4904& BZB  &       &  - & -   & 3 &  -   &     -     &  -   &   -   &   -  &  - &   -  &   -  \\
PKS\,1954$-$388 & 1958.2-3848 & BZQ  & 0.965 & LSP& LSP & 2 &15.30 & 14.3-16.5 & 0.69 & 15.96 &   no & -  &   -  & 1.05 \\
PKS\,2005$-$489 & J2009.5-4850& BZB  & 0.071 & -  & ISP & 1 &12.20 & 11.9-12.5 & 0.22 & 11.35 & dec  & 11 & 0.35 & 0.34 \\
PMN\,J2139$-$4235& 2139.3-4236& BZB  &$>$0.24& -  & HSP & 2 &14.18 & 13.8-14.5 & 0.21 & 14.57 &  no  &  - &  -   & 0.66 \\
PKS\,2155$-$304 & 2158.8-3013 & BZB  & 0.116 &HSP & HSP & 1 &11.93 & 11.4-12.2 & 0.29 & 11.40 & dec  &  7 & 0.49 & 0.43 \\
\hline                                   
\end{tabular}
\end{table*}

For each Group 1 source we report the mean J value (Col. 8),  the peak-to-peak amplitude (Col. 9), the rms deviation of the J dataset with respect to the average J value (Col. 10), and the 2MASS J-band magnitude (Col. 11). We also give in Col. 12 a comment on the general shape of the light curve or on the existence of a long term trend within our monitoring time interval (dec= decreasing; inc=increasing; no= no trend; wav= wavy pattern). Furthermore, we report the  
number of useful observations, the spectral index $\alpha$ from a fit of $F_\nu=A \nu^{-\alpha}$, and the corresponding slope from the 2MASS catalogue values. In these computations we adopted a Galactic absorption according to \citet{Schlegel1998}.
We remark that all the Group 1 sources, save PKS\,0537$-$441, have a lower redshift compared with those in Group 2: this is  expected, since the nearest sources are likely to appear brighter and therefore have a better photometry.

The main characteristics of Group 1 sources can be summarized as follows:

1) Average value: similar to the 2MASS value, except in two cases, PKS\,0447$-$439 and PKS\,2005$-$489.

2) Range: all the sources varied by less than 1 mag, typically 0.6 mag; the larger range showed by PKS\,0537$-$441 is suspect because it is due to only one very discrepant J point with the source on the edge of the field and the corresponding H value is very similar to the remaining values.

3) Long term trend:  clear monotonic trends on a one-year time baseline are apparent in three sources, PKS\,0447$-$439, PKS\,2005$-$489, and PKS\,2155$-$304; this is a subjective impression and depends on the accuracy and number of photometric points. The gamma-ray light curve (see Sect. 3 below) does not always seems consistent with the NIR light curve.

4) The average spectral indexes have a typical error of 0.07 and are generally consistent with the 2MASS values (see Table~\ref{table:1}). The most discrepant cases are PKS\,0537$-$441 and PMN\,J1444$-$3908, but they always lie within 3$\sigma$.

5) Color-flux relation:  a color-flux relation (e.g., bluer-when-brighter, or redder-when-brighter) was not detected for any source; this is expected, given the limited flux variations of all seven sources and the relatively low photometric accuracy (0.05 mag).

None of the seven Group 1 sources underwent a NIR flare during our monitoring. 

For the Group 2 sources we computed the same parameters as Group 1 except for spectral indexes which were derived from their historical magnitudes in the 2MASS catalogue. The results may be summarized as follows:

1) Average value: it is different from the 2MASS value by more than 0.5 mag in  four out of nine cases (PKS\,0208$-$512, PKS\,0700$-$661, PKS\,1424$-$418, PKS\,1954$-$388).

2) Range: most of the sources (seven out of nine) varied by more than 1 mag.

3) Long term trend: only three out of nine sources showed a clear monotonic trend (PMN\,J0334$-$3725, PKS\,0454$-$46, 
PKS\,1424$-$418).

Only one source (PKS\,0208$-$512) showed a flare larger than 1 mag \citep{Blanchard2012}.

For Group 3 sources, all except one (PMN\,J0810$-$7530) are fainter than J=16 in the 2MASS and are therefore expected to be at or beyond the limit of REM. 

\begin{figure}
   \centering
   \includegraphics[width=9cm]{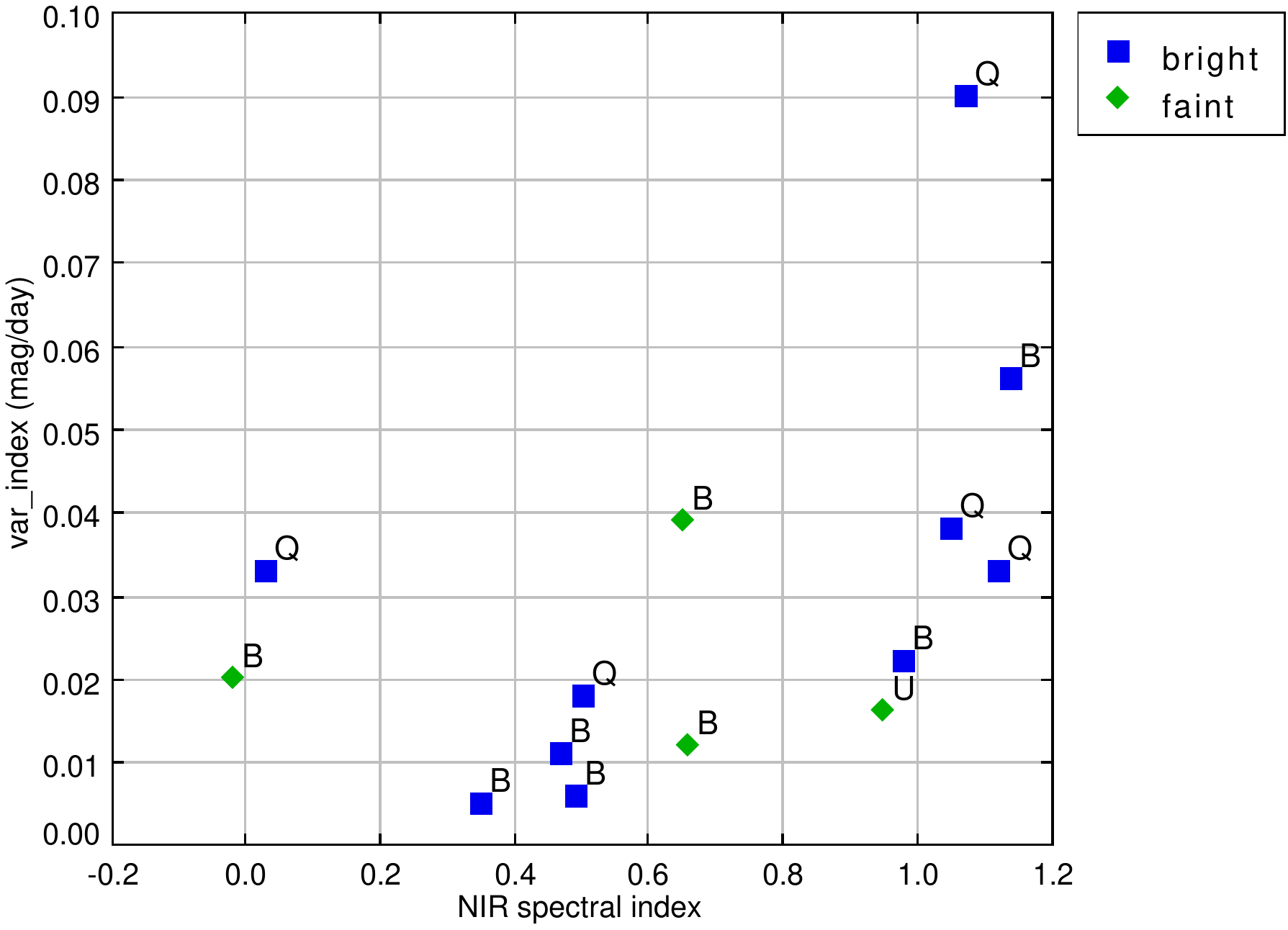}
   \caption{The variability speed index as a function of the NIR spectral index.
   B=BL Lac, Q=QSO, U=unclassified. 
   Squares are objects with absolute J magnitude brighter than -26. Diamonds are fainter sources, including those with unknown redshift.
          }
         \label{fig:speed}
 \end{figure}

Historically, large and fast optical variability was one of the parameters used to define the BL Lac objects (see e.g., Landt et al. 2002).  There is more than one way to define a numerical index to quantify the variability of a source, like those used by Abdo et al. (2010b). Their indexes, however, measure the amplitude, but not the speed of variations. For this purpose we derived a variability-rate index from the light curve of each detected source, computed as the magnitude difference between consecutive points divided by their time interval. The second-fastest value was assumed to be an indicator of the variability speed of each source.  The plot of this index against the NIR spectral index is shown in Fig.~\ref{fig:speed}.

%

It is apparent from this plot that the BL Lac of our sample seem not to be the most variable ones. To get  more insight on this topic we derived the well known $\alpha_{ro}$ and $\alpha_{ox}$  spectral indices \citep{Padovani1995} from the Rome-BZCat  catalogue \citep{Massaro2009} for our detected sources; these indexes are often used to classify the spectral energy distribution of  AGN. From these values then we computed the expected $\nu_{peak}$ value following the Eq. (3) of Abdo et al. (2010) and report the resulting classification as LSP, ISP, or HSP in Col. 8 of Table~\ref{table:1}.  The agreement between our estimate and that reported in the 2LAC catalogue is very good.

\begin{figure}
   \centering
   \includegraphics[width=9cm]{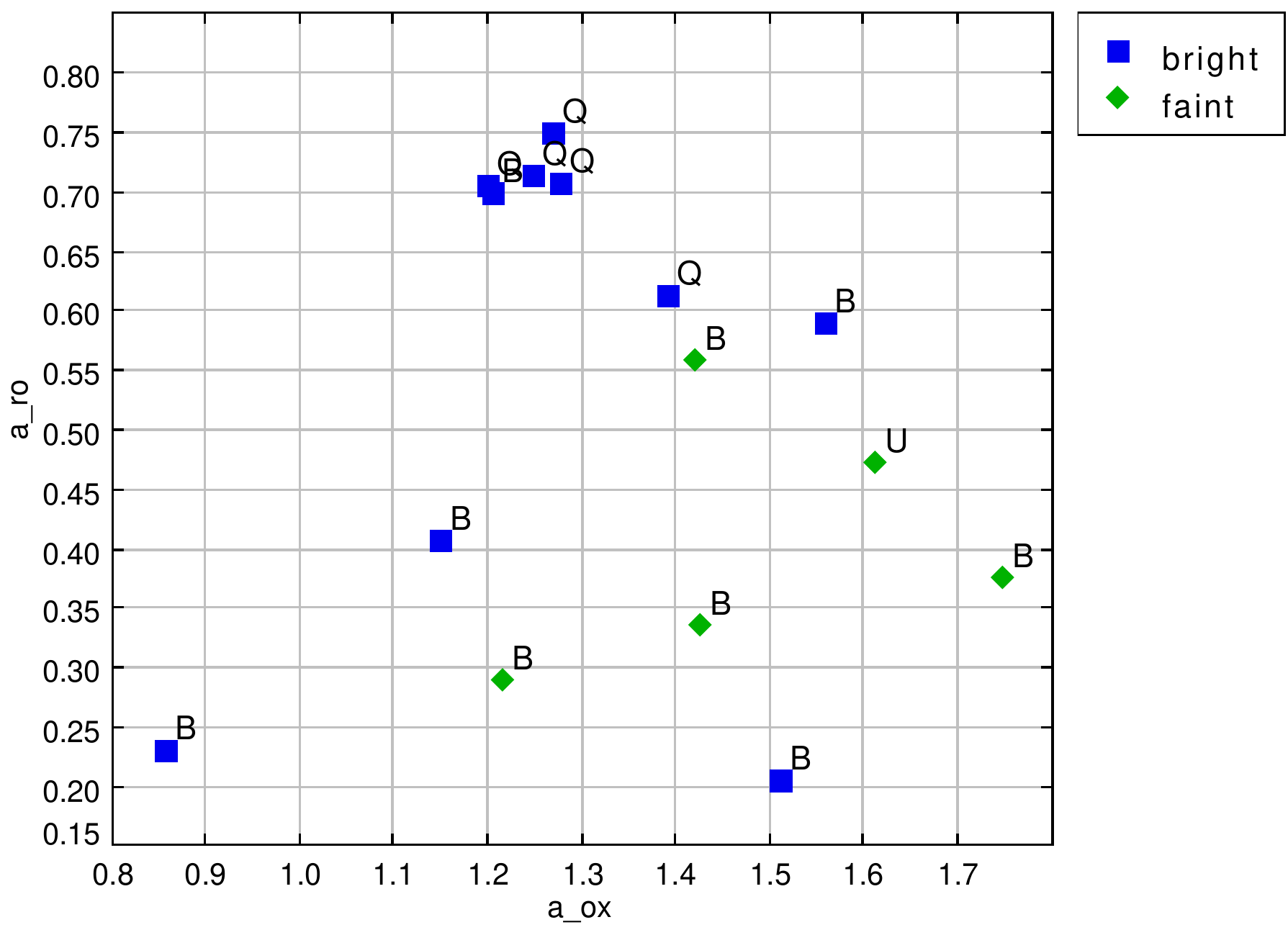}
   \caption{The $\alpha_{ro} - \alpha_{ox}$ plot for the detected sources. Squares are objects with absolute J magnitude brighter than -26. Diamonds are fainter sources, including those with unknown redshift. HSP sources are in the bottom-left corner.
          }
         \label{fig:aroaox}
 \end{figure}

The plot of $\alpha_{ro}$ {\it vs} $\alpha_{ox}$ is shown in Fig.~\ref{fig:aroaox}. There is a trend for the QSOs to lie in the upper-left corner, while BL Lacs are preferentially found in the lower part of the plot. 
In Fig. 12 of \citet{Padovani1995} it is shown that, assuming a synchrotron/inverse-Compton model, blazars move in this diagram, depending on the peak frequency of their synchrotron component, from the upper left (LSP) to the lower÷right along a diagonal line (ISP), and then horizontally toward the left (HSP). Our objects with $\alpha_{ro}$ $\sim$0.3 and $\alpha_{ox} < $1.4 should  therefore have a peak frequency greater than $10^{15}$ Hz, so our NIR observations  sample the rising branch of their synchrotron emission, which is less variable than the falling branch because the radiative cooling time of high÷energy particles is shorter that for the low÷energy ones. 
In the case of QSOs, which have the synchrotron peak at lower frequencies, we are sampling the falling branch of the synchrotron radiation and we therefore expect a somewhat larger variability. We remark also that in our sample QSOs have a value of $\alpha_{ro}> 0.65$, so that this index is a good discriminant between QSOs and BL Lacs. 

Finally, we show the scatter plot between the $\alpha_{ro}$ and the NIR spectral index (Fig. 4): we found a good correlation (r=0.80)  for our BL Lac sources, while the QSOs show no correlation. The most likely explanation is that in the case of BL Lacs the SED in the radio-NIR regions is essentially produced by the same synchrotron component and is smooth, while  for QSOs the SED has a local minimum at about 1 $\mu$m, with individual objects showing large differences (see e.g., the pioneering work by Elvis et al. 1994).

%
%
 \begin{figure}
   \centering
   \includegraphics[width=9cm]{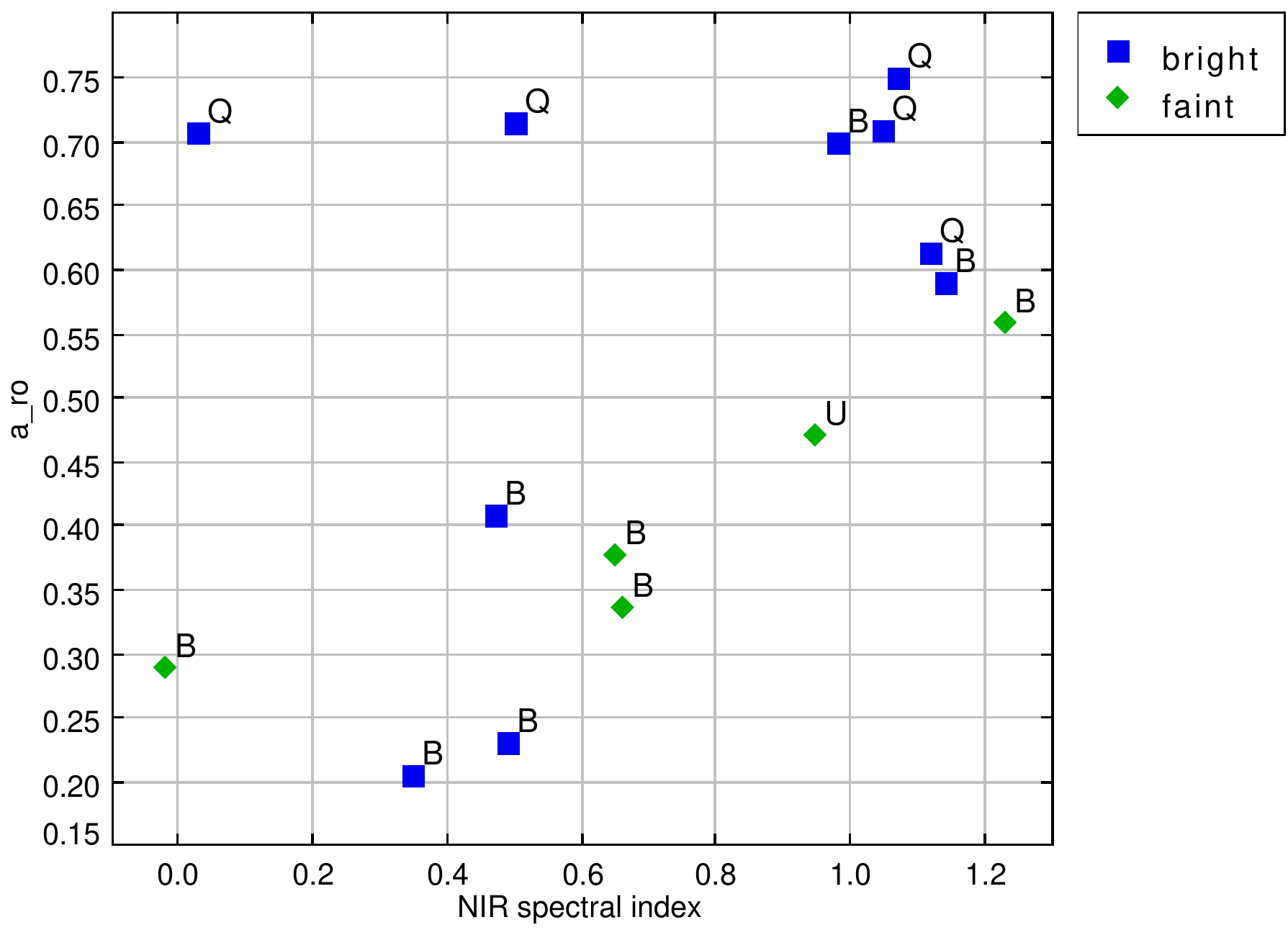}
   \caption{The  $\alpha_{ro}$  vs NIR spectral index for the Group 1 and 2 sources. Squares are objects with absolute J magnitude brighter than -26. Diamonds are fainter sources, including those with unknown redshift.
          }
         \label{fig:speedaro}
 \end{figure}

\section{Gamma-ray data}
\label{sect:gammaraydata}
The Large Area Telescope (LAT) instrument of the \textit{Fermi} Gamma-ray Space Telescope \citep{Atwood2009} observes the whole sky every three hours. Thus, we were able to generate gamma-ray light curves of these TANAMI-REM sources from \textit{Fermi}/LAT data.
 
Light curves  were created for all the sources with one week time bins. We applied the binned likelihood method implemented in the standard LAT Science Tools\footnote{http://fermi.gsfc.nasa.gov/ssc/data/analysis/scitools/overview.html} (version v9r28p0). We analyzed data collected, in the 100 MeV - 100 GeV energy range, during 2011. We analyzed Pass 7 {\it Source} event class data (with instrument response functions version P7SOURCE\_V6\footnote{http://www.slac.stanford.edu/exp/glast/groups/canda/lat\_Performance.htm}) in a circular region of interest (ROI) of 10$^\circ$ radius, centered at the position of the  source. 
Standard event cuts on the zenith angle ($<$ 100$^\circ$ ) and on the rocking angle ($<$ 52$^\circ$ ) to limit contamination from the Earth limb were also applied. 
The ROI model used by the binned likelihood method includes all point sources from the 2FGL catalog (Nolan et al. 2012) located within 15$^\circ$ from the  source. In the model, only sources located within a 3$^\circ$  radius centered on the source position had parameters left free to vary; sources between 15$^\circ$ and 3$^\circ$ had their parameters set to the value of the 2FGL catalog.  The source of interest  was modeled as a power law with both normalization and spectral index free to vary.  Standard diffuse Galactic and isotropic components were also included in the model\footnote{http://fermi.gsfc.nasa.gov/ssc/data/access/lat/BackgroundModels.html}. The normalization parameters of the diffuse models were left free in the fit.
All our sources were rather faint, with a median flux below 5$\times$10$^{-7} $phot~cm$^{-2}$~s$^{-1}$, and nine of them below 1.5$\times$10$^{-7}$ phot~cm$^{-2}$~s$^{-1}$ . In several cases only upper limits could be derived.

\begin{figure}
   \centering
   \includegraphics[width=9cm]{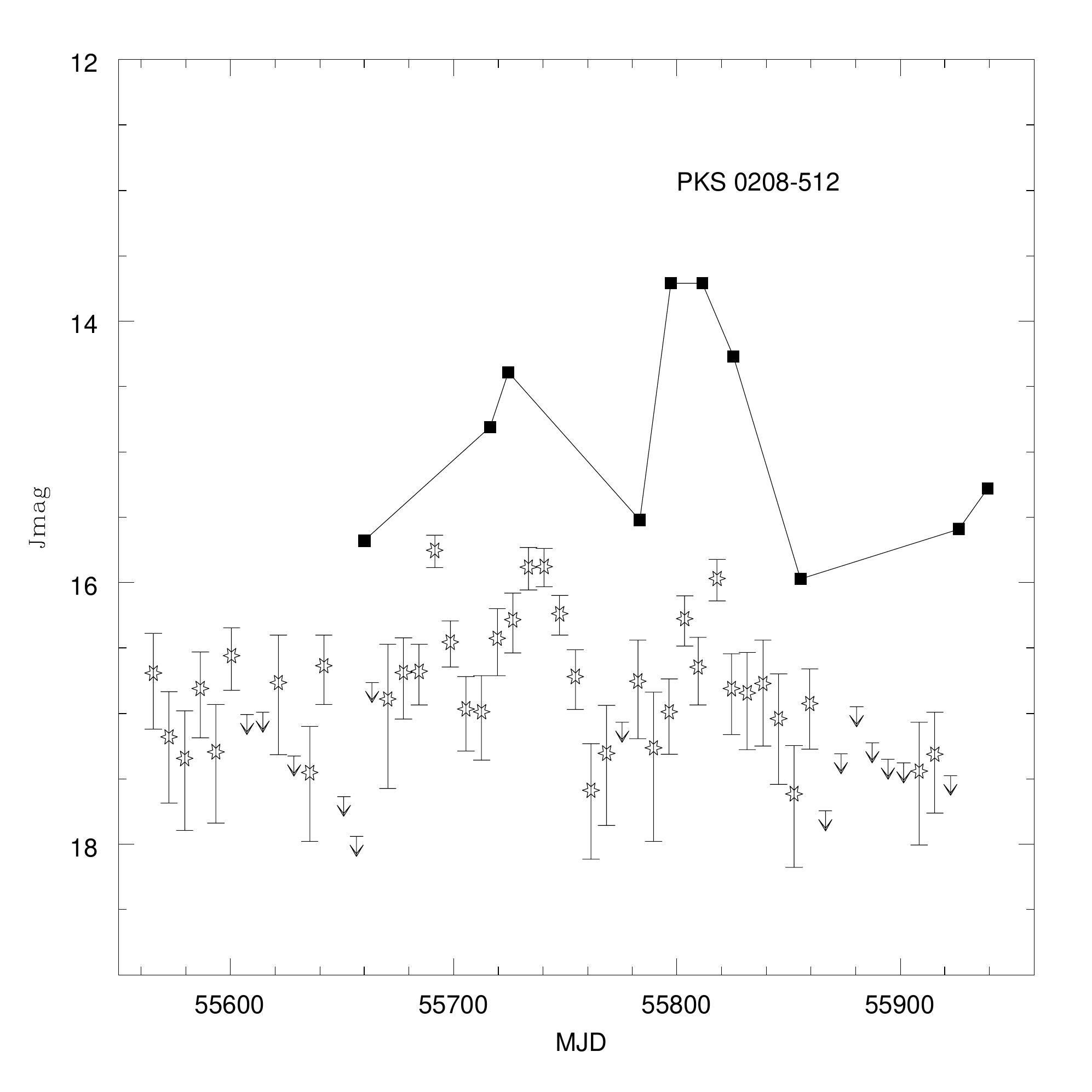}
   \caption{The J-band light curve (filled squares) and gamma-ray light curve (stars) of PKS\,0208$-$512. Upper limits are shown as arrows. The abscissa is in MJD and the ordinate in magnitudes. The gamma-ray count rate has been converted into magnitudes as -2.5 log[ F/(phot~cm$^{-2}$~ s$^{-1}$)]; a value of 20 corresponds to 1$\times$10$^{-8}$ phot~cm$^{-2}$ s$^{-1}$. 
          }
         \label{fig:0208}
 \end{figure}

A correlation between NIR and gamma-ray fluxes was found by Chatterjee et al. (2012) and Bonning et al. (2012) from a blazar subsample of the Yale/SMARTS monitoring program. We looked for a similar correlation in our data with a Kendall rank test, (Akritas \& Siebert 1996) using the so-called ASURV bhkmethod implemented in iraf/stsdas. For three sources we also used the kendalltau-test as provided in the python/numpy/statistics package. 
The probability of a chance correlation for four sources is small, i.e., PKS\,0537$-$441(0.019), PKS\,0521$-$360 (0.063), PKS\,2155$-$304 (0.061),  and PKS\,1424$-$418 (0.128), while for the others no conclusion can be drawn.
As an example of the typical data quality we plot in Fig.~\ref{fig:0208} the J-band and gamma-ray light curves of PKS\,0208-512, the only source showing a flare during our monitoring. 

Our NIR and Gamma-ray monitoring show a sure, albeit small, variability for the radio galaxy PKS\,0625$-$354, also in favor of a BL Lac classification. The TANAMI VLBI images at 8.4 GHz (Ojha et al. 2010) show a one-sided parsec-scale jet, indicative of substantial Doppler boosting. Multi-epoch observations to study the jet kinematics are in progress and will help to constrain the Doppler factor and the angle to the line of sight. We recall that Wills et al. (2004) obtained a spectrum of this source with a marked UV excess, suggesting a BL Lac nature.

For each of our Group 1 and Group 2 sources we computed the average gamma-ray/J-band flux ratio, which can be considered an approximate estimate of the ratio between Compton and synchrotron total power emission (Compton dominance): we found a wide range of values (about a factor of 30). If this flux ratio is plotted against the $\nu_{peak}$ index, computed as in Abdo et al. 2010, a nice correlation is apparent (see Fig. 5) with correlation factor r=-0.85. The BL Lac objects are well separated from the QSOs, suggesting the existence of a physical sequence: objects with lower synchrotron peak frequency (QSOs) apparently have a higher gamma-ray/NIR flux ratio.
This is in agreement with the results from the 2LAC sample of {\it Fermi}-detected blazars extensively discussed by D'Abrusco et al (2012) (see their Fig.~8).

We also looked for a correlation between our NIR spectral index and the gamma-ray index, as
D'Abrusco et al. (2012) did, but no clear trend was apparent, probably because of the small number of sources and the rather large uncertainties of our gamma-ray spectral indexes.

\begin{figure}
   \centering
   \includegraphics[width=8cm]{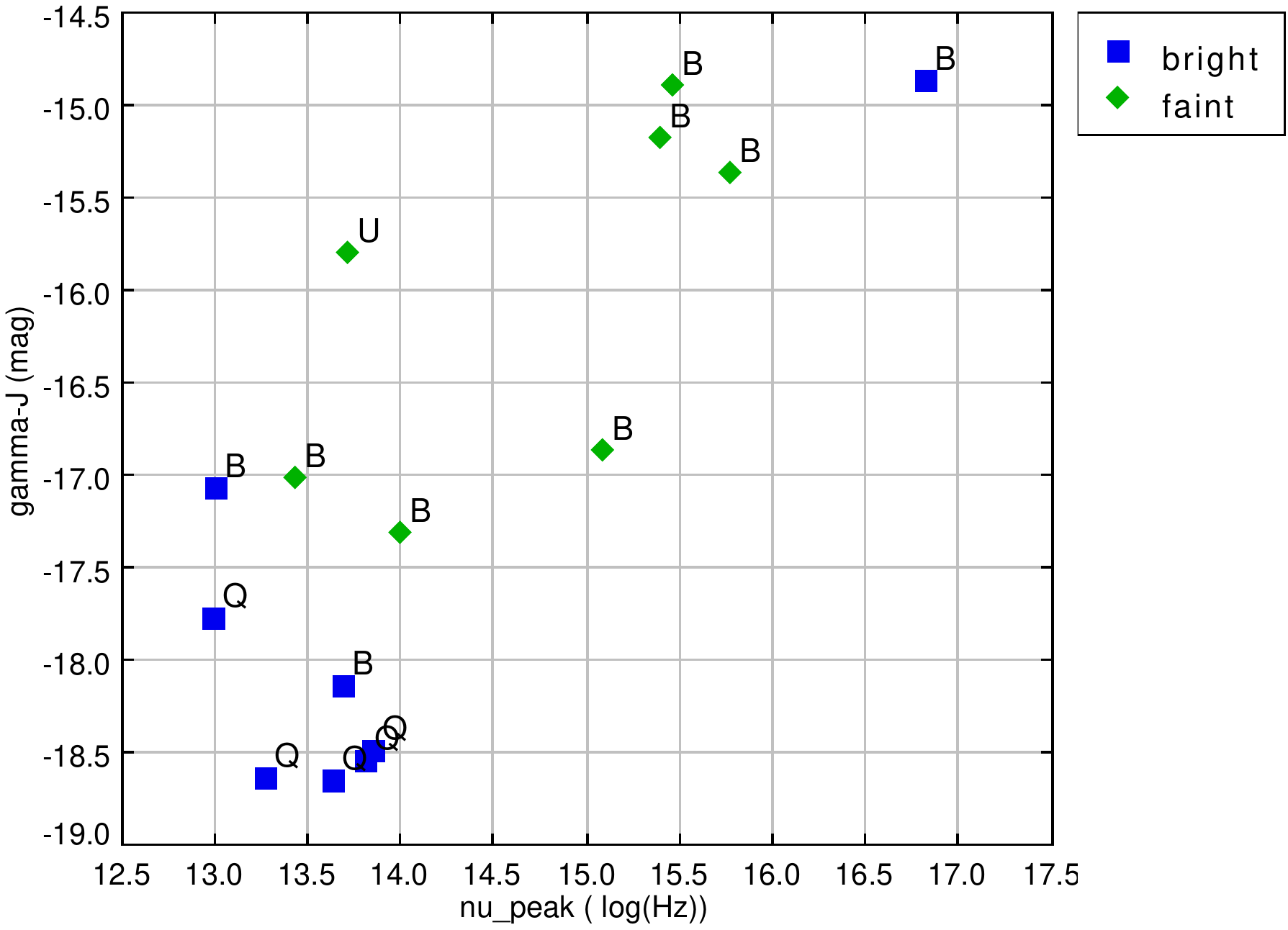}
   \caption{Ratio of gamma-ray/J-band average flux during our monitoring (in arbitrary magnitude units) against the $\nu_{peak}$ index. Sources in the upper-right corner (BL Lacs) have lower gamma/NIR flux ratio, i.e., are relatively gamma-ray faint.
          }
         \label{fig:gamir}
 \end{figure}
\section{Conclusions}
\label{sect:conclude}
We performed a one-year J, H, K monitoring of 22 blazars from the sample of the TANAMI radio-monitoring project. Sixteen of them were always detected and light curves in the J band derived.  For the brightest ones, the R-band magnitudes were also measured and found consistent with the J values.
All sources showed some variability, always larger than 0.5 mag, which was on average larger for the QSOs. In our opinion this may be due to the fact that most of our BL Lac sources are of the HSP-type and therefore expected to be less variable in the NIR range.
For the seven brightest sources we could also derive meaningful NIR spectral indexes, which were nearly always found in agreement with the values derived from the 2MASS catalogue. No appreciable changes in the spectral indexes were detected, but they were not expected to be found given that no large variations in flux densities occurred during our monitoring.

The variability in the NIR and gamma-ray bands was fairly well correlated only in four sources, as expected in the classical scenario where synchrotron/inverse-Compton is the main emission process in blazars. This is in agreement with the results of the 2LAC sample of {\it Fermi}-detected blazars \citep{Ackermann2011}. For most sources we found no evidence of correlation, likely because of the poor NIR sampling and the gamma-ray faintness of the sources. 

Finally, we found that the well-known $\alpha_{ro}$ parameter, which is a good discriminator between BL Lac and QSO, also appears to be well correlated with the gamma-ray/NIR flux ratio. 
 
\begin{acknowledgements}

This research was funded in part by NASA through  Fermi Guest Investigator grant NNH09ZDA001N (proposal number 31263) and grant NNH10ZDA001N (proposal number 41213). This research was
supported by an appointment to the NASA Postdoctoral Program at the
Goddard Space Flight Center, administered by Oak Ridge Associated
Universities through a contract with NASA. This research has made use of the 
SIMBAD database, operated at CDS,
Strasbourg, France, and also of the Two Micron All-Sky Survey
database, which is a joint project of the University of Massachusetts and
the Infrared Processing and Analysis Center/California Institute of
Technology. This research has made use of NASA's Astrophysics Data 
System Bibliographic Services.
      
The \textit{Fermi} LAT Collaboration acknowledges generous ongoing support
from a number of agencies and institutes that have supported both the
development and the operation of the LAT as well as scientific data analysis.
These include the National Aeronautics and Space Administration and the
Department of Energy in the United States, the Commissariat \`a l'Energie Atomique
and the Centre National de la Recherche Scientifique / Institut National de Physique
Nucl\'eaire et de Physique des Particules in France, the Agenzia Spaziale Italiana
and the Istituto Nazionale di Fisica Nucleare in Italy, the Ministry of Education,
Culture, Sports, Science and Technology (MEXT), High Energy Accelerator Research
Organization (KEK), and Japan Aerospace Exploration Agency (JAXA) in Japan, and
the K.~A.~Wallenberg Foundation, the Swedish Research Council, and the
Swedish National Space Board in Sweden.

Additional support for science analysis during the operations phase is gratefully
acknowledged from the Istituto Nazionale di Astrofisica in Italy and the Centre National d'\'Etudes Spatiales in France.     
\end{acknowledgements}

\newpage

\end{document}